\documentclass[twocolumn,aps,prb,amsmath,showpacs]{revtex4-1}

\pdfoutput=1

\usepackage[utf8]{inputenc} 
\usepackage[T1]{fontenc} 
\usepackage{microtype}
\usepackage{mathrsfs}
\usepackage{amsmath}
\usepackage{physics}
\usepackage{balance}
\usepackage{natbib}
\usepackage{adjustbox}

\usepackage{graphicx}
\makeatletter
\def\@biblabel#1{[\textbf{#1}]}
\makeatother
\usepackage[dvipsnames]{xcolor}

\usepackage{dcolumn}
\usepackage{bm}

\usepackage[colorlinks = true,
            linkcolor = blue,
            urlcolor  = blue,
            citecolor =red,
            anchorcolor = blue]{hyperref}
\usepackage{verbatim}
\usepackage{color,ulem}
\usepackage[english]{babel}

\usepackage[scr]{rsfso}

\usepackage{blindtext}

\begin{document}

\preprint{IFT-UAM/CSIC-20-55}

\title{\Large Soft mode theory of ferroelectric phase transitions in the low-temperature phase}

\author{Luigi Casella}%
\affiliation{
Department of Physics "A. Pontremoli", University of Milan, via Celoria 16, 20133 Milan, Italy.
}

\author{Alessio Zaccone}%
 \email{alessio.zaccone@unimi.it }
\affiliation{
Department of Physics "A. Pontremoli", University of Milan, via Celoria 16, 20133 Milan, Italy.\\
Cavendish Laboratory, University of Cambridge, JJ Thomson
Avenue, CB30HE Cambridge, U.K.
}

\begin{abstract}
Historically, the soft mode theory of ferroelectric phase transitions has been developed for the high-temperature (paraelectric) phase, where the phonon mode softens upon decreasing the temperature. In the low-temperature ferroelectric phase, a similar phonon softening occurs, also leading to a bosonic condensation of the frozen-in mode at the transition, but in this case the phonon softening occurs upon increasing the temperature.
Here we present a soft mode theory of ferroelectric and displacive phase transitions by describing what happens in the low-temperature phase in terms of phonon softening and instability. 
A new derivation of the generalized Lyddane-Sachs-Teller (LST) relation for materials with strong anharmonic phonon damping is also presented which leads to the expression $\varepsilon_{0}/\varepsilon_{\infty}=|\omega_{LO}|^{2}/|\omega_{TO}|^{2}$. 
The theory provides a microscopic expression for $T_c$ as a function of physical parameters, including the mode specific Gr\"uneisen parameter. The theory also shows that $\omega_{TO} \sim (T_{c}-T)^{1/2}$, and again specifies the prefactors in terms of Gr\"uneisen parameter and fundamental physical constants. Using the generalized LST relation, the softening of the TO mode leads to the divergence of $\epsilon_0$ and to a polarization catastrophe at $T_c$. A quantitative microscopic form of the Curie-Weiss law is derived with prefactors that depend on microscopic physical parameters. 
\end{abstract}

\maketitle
\section{Introduction}
Anderson and Cochran, around 1960, independently predicted that ferroelectric phase transitions originate from a soft mode, i.e. from lattice dynamical instability~\cite{Cochran_1959,Anderson}, although the same prediction appeared (in Russian) some 10 years earlier by Vitaly Ginzburg, see Ref.~\cite{Ginzburg_2001} for a historical perspective. 
Those seminal works were concerned with the ferroelectric phase transition approached from above, i.e. with the paraeletric to ferroelectric phase transition. 
In practice, the Green's function of a phonon which undergoes softening as the temperature in the high-symmetry phase is lowered, can be analyzed starting from the most generic phonon propagator:
\begin{equation}
[\Omega^{2}-\omega_{0}^{2}-2\omega_{0}(\Delta - i \Gamma)]^{-1} \nonumber
\end{equation}
where, using the standard notation for Green functions, $\Omega$ is the frequency in the argument of the Green function, $\omega_{0}$ is the "bare" phonon frequency,  $\Delta$ and $\Gamma$ are the anharmonic parameters associated with shift or renormalization of the phonon frequency and phonon linewidth/damping, respectively. These parameters can be computed using the self-consistent phonon (SCP) theory, originally developed by Hooton~\cite{Hooton} and later formalized in the language of many-body theory~\cite{Tadano}.
The response function is peaked at a frequency $\omega\approx \omega_{0}^{2} + 2\omega_{0}\Delta$.
The frequency renormalization parameter $\Delta$ is proportional  to the occupation numbers of the final states into which  the phonon is going to decay. For cubic anharmonicity and considering only the lowest-order contributions, one has $\Delta \sim (n_{1} + n_{2} +1)$, if the phonon decays into two phonons with occupation numbers $n_{1}$ and $n_{2}$, or $\Delta \sim (n_{1} - n_{2})$ for scattering with absorption of a phonon~\cite{Tadano}. If, instead, one considers only quartic interactions, as this is often the dominant contribution to the phonon renormalization\cite{Dove}, we have, to leading order~\cite{Tadano}, $\Delta \sim (2n_{2} +1)$.
In all cases, with $n_{1,2}=[\exp{\hbar \omega/k_{B}T} -1]^{-1}$, at high temperature $\hbar \omega \ll k_{B}T$, the final result for $\Delta$ will be:
$\Delta=2^{-1} \omega_{0}^{-1}  C T$, where $C$ is a constant. This result is confirmed by more quantitative calculations of the self-consistent equations. This in turn leads to
\begin{equation}
\omega^{2} = C (T-T_{c})   
\label{paraelectric}
\end{equation}
in agreement with what is expected from Landau theory~\cite{Ginzburg_2001}. The same result for the soft mode instability and the soft mode dependence on $T$ in the paraelectric phase can be obtained based on microscopic phonon physics\cite{Cowley_book} and also, with a different approach, from microscopic Hamiltonians for the local lattice distortions\cite{Pytte}.

The above description, based on the phonon frequency renormalization due to (weak) anharmonicity, is the basis of soft mode theory in the high-temperature (paraelectric) phase.
Historically, however, less attention has been paid to the soft mode instability in the low-temperature ferroelectric phase. Indeed, if the phase transition coincides with the condensation of a soft phonon upon decreasing $T$ from above in the paraelectric phase, a similar process must occur upon increasing $T$ in the ferroelectric phase\cite{Venkataraman}. This is because the frozen-in distortion gives rise to a vibrational mode which attempts to restore the higher symmetry lost upon going from paraelectric to ferroelectric. Hence, a soft mode exists, as is well known for all displacive ferroelectrics, also in the ferroelectric phase. This soft mode becomes unstable at the phase transition, although its temperature dependence cannot be given by Eq.\eqref{paraelectric}, because the frequency of this mode decays upon increasing $T$, which is of course the opposite of what happens in the high-$T$ phase. From experiments and previous phenomenological work it is known that this mode dies off at the transition as $\omega^{2} = C (T_{c}-T)$, although microscopic derivations of this result are not available. 

In this contribution, we provide a microscopic theory of the soft mode in the low-$T$ ferroelectric phase, which leads to the derivation of the Curie-Weiss law upon approaching the ferroelectric to paraeletric transition from below. The theory provides also prefactors in terms of physical parameters related to the microscopic phonon physics and fundamental physical constants.

We start from deriving a generalized Lyddane-Sachs-Teller (LST) relation which accounts for strong anharmonic damping. We then consider the temperature evolution of an optical phonon subject to strong anharmonicity and show that it eventually becomes unstable at the critical temperature $T_c$. We then show that this result, implemented in the LST relation with damping, leads to the Curie-Weiss law. 
These results show that, contrary to what happens in the high-$T$ phase, where the strength of anharmonicity and the damping are not essential~\cite{Cowley_book}, the soft mode instability in the paraelectric phase is instead driven by strong anharmonic damping of the optical phonon, which softens upon increasing the temperature towards $T_c$.

\section{Approximate generalized LST relation from previous works}
As is customary, one works with a harmonic equation of motion for the relative displacement field, with an additional term that describes the effect of the electric field on the charges. This equation must be coupled with the equation for the polarization, which has a term proportional to the relative displacement field and a term linear in the electric field. Assuming a linear polarizability of the charged atoms, this leads to~\cite{Born_lattice}

\begin{equation}
    \begin{cases}
    \Ddot{\Vec{u}}=b_{11}\Vec{u}+b_{12}\Vec{E} \\
    \Vec{P}=b_{21}\Vec{u}+b_{22}\Vec{E}.
    \end{cases}
    \label{dis&pola}
\end{equation}
Analytical expressions for the $b$ coefficients of Eq.\eqref{dis&pola} were found by Huang in his pioneering work on the polariton\cite{Huang-1950,Huang-1951}, and can be found also in the monograph by Born and Huang~\cite{Born_lattice}.
To make the comparison with these works easier, we will use the same notation adopted by Born and Huang~\cite{Born_lattice} throughout the manuscript. 
 
With the usual definition of the dielectric function ($\vec{E}\varepsilon(\omega)=\vec{E}+4\pi\vec{P}$, assuming $\varepsilon(\omega)$ to be a scalar function), we can solve Eq.\eqref{dis&pola}. Since we have no free charges, the divergence of the electric displacement is zero. 
The mathematical derivation of the solution will be reported later in this work for a more general case. 
We find three solutions for Eq.\ref{dis&pola}, one longitudinal and two transverse, with equal frequencies. The relation between the frequency of the two phonons is: 
\begin{equation}
    \label{LST}
    \omega^{2}_{LO}=\dfrac{\varepsilon_{ 0}}{\varepsilon_{\infty}} \omega_{TO}^{2}
\end{equation}
and it is known as the Lyddane-Sachs-Teller relation (LST), and it was first found in their famous paper \cite{LSToriginal}. 
This relation holds for oscillations that are not affected by damping. In the LST model, the optical phonons are normal modes of oscillation with infinite lifetime. Whenever a damping effect is present, we expect the lifetime of the particle to be of finite lifetime. Mathematically, this is described by giving an imaginary part to the frequency of the quasi-particle, that is equal to the inverse of its characteristic lifetime. This finite-lifetime modes are often referred to as \textit{quasi-normal modes}. 

We are interested in deriving a LST relation that holds also for quasinormal modes. One possible way to obtain such relation was introduced by Barker~\cite{Barker}. This derivation relies on definitions of the transverse and longitudinal phonon frequencies, based on their role in the dielectric function. The longitudinal phonon frequency is defined as the frequency for which the dielectric function equals zero. The longitudinal phonon is assumed to be the frequency for which the imaginary part of the dielectric function peaks. This is based on the assumption of a long lifetime of the particle, so that the dielectric function does not deviate significantly from the undamped case. 
With these assumptions, Barker obtained an expression for the longitudinal phonon frequency:
\begin{equation}
    \label{barker wl}
    \omega_{LO}=-i\dfrac{\gamma}{2}\pm \sqrt{\omega^{2}_{TO}\dfrac{\varepsilon_{0}}{\varepsilon_{\infty}}-\dfrac{\gamma^{2}}{4}}
\end{equation}
where $\varepsilon_{0}$ and $\varepsilon_{\infty}$ are the limits for vanishing or infinite frequency of the dielectric function, respectively, and $\gamma$ is the damping of the Lorentz  oscillator model from which the dielectric function is derived.

Obviously, taking the absolute value of Eq.\eqref{barker wl}, we can obtain a relation between the two frequencies:
\begin{equation}
    \label{barker LST}
     \dfrac{\varepsilon_{0}}{\varepsilon_{\infty}}=\dfrac{|\omega_{LO}|^{2}}{\omega_{TO}^{2}}.
\end{equation}

This approach has two main limitations to its generality. First of all, the transverse frequency is assumed to be real, i.e. an oscillation with infinite lifetime. Also, this result is limited to systems with small damping since it assumes that the dielectric function must have both a sharp peak and a frequency range in which it is negative. For systems with strong anharmonicity, these assumptions do not hold.  Furthermore, the effect of damping is considered in the Barker's LST expression derived above, only for the longitudinal modes.
These limitations pose a huge problem for developing a soft mode theory of ferroelectric phase transitions in the ferroelectric phase, where the softening mode is generally a transverse (TO) mode~\cite{Scott}. In the next section we amend this problem by deriving a more general LST relation, which allows us to account for anharmonic damping of the TO mode, and which is more generally valid for systems with strong anharmonic damping.

\vspace{0.2cm}

\section{A more rigorous generalized LST}
We derive a more general LST relation with an effective field theory approach. We assume that all the damping effects on the waves can be described by an additional term in the first line of Eq.\eqref{dis&pola} that breaks the time inversion symmetry. We assume that such a term can be written as a function that depends on the time derivative of the displacement field,

\begin{equation}
    \begin{cases}
    \Ddot{\Vec{u}}=b_{11}\Vec{u}+ f(\dot{\Vec{u}}) +b_{12}\Vec{E} \\
    \Vec{P}=b_{21}\Vec{u}+b_{22}\Vec{E}.
    \end{cases}
    \label{Damped_dis&pola}
\end{equation}

Also, since there are no free charges in the system, the divergence of the electric displacement $\vec{D}$ equals zero. From its definition, a relation between the electric field and the polarization can be found,
\begin{equation}
    \begin{split}
        &\nabla\cdot\vec{D}=\nabla\cdot\left(\vec{E}+4\pi\vec{P}\right) = 0\\&\nabla\cdot\vec{E} = -4\pi\nabla\cdot\vec{P}.
    \end{split}
\end{equation}
Substituting in the expression for the polarization of Eq.\eqref{Damped_dis&pola}, we obtain:
\begin{equation}
\label{divergenceEu}
\begin{split}
    &\nabla\cdot\vec{E} = -4\pi\nabla\cdot\big(b_{21}\Vec{u}+b_{22}\Vec{E}\big) \\& \nabla\cdot\vec{E} = -\frac{4\pi b_{21}}{1+4\pi b_{22}}\nabla\cdot\Vec{u},
\end{split}
\end{equation}
where the divergence of the displacement field is equal to the divergence of the irrotational part of the field, which is the longitudinal component. This fact is consistent with the Helmholtz theorem\cite{Jackson}. Integrating Eq.\eqref{divergenceEu}, we obtain a relation for the longitudinal displacement field and the electric field, as follows:
\begin{equation}
\label{relation: electric - longitudinal displacement}
    \vec{E} = -\frac{4\pi b_{21}}{1+4\pi b_{22}}\vec{u}_{L}.
\end{equation}
The latter result, replaced in the first line of Eq.\eqref{Damped_dis&pola}, yields an equation for the displacement field,
\begin{equation}
    \Ddot{\Vec{u}}=b_{11}\Vec{u}+ f(\dot{\Vec{u}}) -\frac{4\pi b_{12}b_{21}}{1+4\pi b_{22}}\vec{u}_{L}.
\end{equation}
If we assume the damping term to be of the standard Langevin type, $f(\dot{\Vec{u}})=\Gamma\dot{\Vec{u}}$, and recalling that the direction of the polarization of the three waves forms a basis for the space of the displacement, we obtain two different equations for the longitudinal and the transverse waves,
\begin{equation}
\begin{split}
    &\Ddot{\Vec{u}}_{L}=\left(b_{11}-\frac{4\pi b_{12}b_{21}}{1+4\pi b_{22}}\right)\Vec{u}_{L}+ \Gamma\dot{\Vec{u}}_{T}   \\
    &\Ddot{\Vec{u}}_{T}=b_{11}\Vec{u}_{T}+ \Gamma\dot{\Vec{u}}_{T}.
\end{split}
\end{equation}

Upon Fourier transforming the above expressions we obtain: 
\begin{equation}
\label{LoToeq}
\begin{split}
    &\omega^{2}_{LO}=-\left(b_{11}-\frac{4\pi b_{12}b_{21}}{1+4\pi b_{22}}\right)+i\omega_{LO}\Gamma\\
    &\omega^{2}_{TO}=-b_{11}+i\omega_{TO}\Gamma.
\end{split}
\end{equation}
Using the expressions found by Huang\cite{Huang-1950,Huang-1951} for the coefficients $b_{ij}$,  the roots of Eq.\eqref{LoToeq} are easily obtained
\begin{equation}
\label{general_LOTO}
    \begin{split}
               & \omega_{LO}= -i\dfrac{\Gamma}{2} \pm\sqrt{\omega_{0}^{2}\dfrac{\varepsilon_{0}} {\varepsilon_{\infty}} -\dfrac{\Gamma^{2}}{4}}   \\&
         \omega_{TO}= -i\dfrac{\Gamma}{2}\pm\sqrt{\omega_{0}^{2}-\dfrac{\Gamma^{2}}{4}}.  \\ 
    \end{split}
\end{equation}
Taking the absolute value of each complex frequency we obtain a new generalized LST relation with anharmonic damping,
\begin{equation}
    \label{barker LST}
     \dfrac{\varepsilon_{0}}{\varepsilon_{\infty}}=\dfrac{|\omega_{LO}|^{2}}{|\omega_{TO}|^{2}}.
\end{equation}
This relation give us a double advantage. First it allows us to take into account the damping effects on the transverse wave, which was not the case for the Barker expression, Eq. (5). Also, with this new expression we are free to use any kind of damping model and since the damping term was a linear function of the field we can use different damping models (or different parameters for the same model) on each wave. 
The relation is easily extended to the co-existence of many modes by taking the product over the modes in both the numerator and denominator of the left-hand side, as is customary~\cite{Fleury}.

With respect to previously derived generalized LST relations~\cite{Barker,Changdoi:10.1002/pssb.19680280224}, Eq. (14) has the advantage of compactly accounting for the effect of anharmonic damping on the TO modes, which is the crucial mode that undergoes softening at the ferroelectric transition upon coming from the low-$T$ phase. Moreover, this equation compactly expresses the LST relation for  \textit{quasi normal} modes, i.e. it simultaneously takes into account both the particle frequency and its lifetime in a compact way. In the following, we will use the above LST relation to describe the polarization catastrophe at the ferroelectric transition as approached from low-temperature, and to derive the Curie-Weiss law in the ferroelectric phase.

\section{Soft mode instability in the low-$T$ phase}
The derivation presented above may allow us to make more accurate predictions of structural transitions caused by the softening of the optical modes in the low-temperature phase. The softening of a phonon is the process in which the energy of a mode goes to zero. This frozen-in zero-energy mode is connected with the loss of the restoring forces of the lattice on the displacement field in certain directions. Such process leads eventually to a structural transition~\cite{Dove}.   
The transverse mode in a solid that displays polar vibrations is always lower in energy than the longitudinal modes~\cite{Dove}, see also the section below. If we want to study structural transitions induced by the softening of optical modes, the best candidate should be the energetically lower mode, i.e. the transverse mode. Therefore, in this context, the usefulness of the generalized LST relation derived in the previous section will prove crucial.

\subsection{Optical phonon softening from Klemens damping}. We model the optical phonon softening upon increasing the temperature in the low-$T$ phase by accounting for the microscopic physics of the anharmonic decay process of the optical phonon. Klemens, in a famous paper \cite{Klemens}, calculated the rate of decay of an optical phonon into two acoustic ones via a Boltzmann-type master kinetic equation and perturbation theory. The three-phonon process is the dominant process under standard conditions, unless one operates at temperatures much higher than room temperature (where higher-order processes become more important)~\cite{Balkanski}. In spite of the approximations used, the Klemens damping model still provides a reasonably accurate estimate of the damping coefficient in comparison with well controlled experimental measurements~\cite{Kitajima}.

This process, governed by anharmonic interactions, is considered the dominant decay process for optical phonons.  The result of Klemens' estimate yields an expression for the mean lifetime of the particle:
\begin{equation}
    \label{klemenstau}
        \dfrac{1}{\tau}=\omega\dfrac{J}{24\pi}\gamma^{2}_{G}\dfrac{\hbar\omega}{M v^{2}}\dfrac{a^{3}\omega^{2}}{v^{3}}C(\alpha,\beta)\left[1+\dfrac{2}{e^{x}-1}\right]  
\end{equation}
with
\begin{equation}
C(\alpha,\beta)=\dfrac{2}{\sqrt{3}}\dfrac{\alpha-\beta}{\alpha+\beta};\,\,\,x=\dfrac{\hbar\omega}{2k_{B}T}.
\end{equation}
Here, $\omega$ is the frequency of the optical phonon that decays via the anharmonic process, which could be either longitudinal (LO) or transverse (TO). More precisely, $\omega$ is the frequency in the limit of zero temperature. Furthermore, $J$ is an integer number that counts the allowed branch transitions from the optical mode to the acoustic ones,  $\gamma_{G}$ is the Gr\"uneisen parameter of the lattice, $a^{3}$ is the volume per atom, $M$ is the reduced atomic mass, and $v$ is the speed of the acoustic phonons in the Debye approximation, while $C$ is a correction coefficient of the order $\mathcal{O}(0.1)$, which, in Klemens' original derivation, depends on the two spring constants, $\alpha$ and $\beta$, of a prototypical alkali halide-type lattice. 
For the rest of the paper we will summarize all the physical prefactors in Eq.\eqref{klemenstau} using the single parameter $\zeta$, as  follows,
\begin{equation}
    \label{klemens_summarized}
    \dfrac{1}{\tau}=\omega^{5}\zeta\left[1+\dfrac{2}{e^{x}-1}\right],
\end{equation}
with 
\begin{equation}
\zeta \equiv \dfrac{J}{24\pi}\gamma^{2}_{G}\dfrac{\hbar}{M v^{2}}\dfrac{a^{3}}{v^{3}}C(\alpha,\beta).
\end{equation}

Since the imaginary part of the damped (quasi normal) mode is the inverse of the typical lifetime of the particle, from Eq.\eqref{general_LOTO} we have that:

\begin{equation}
 \label{Gamma_tau}
    \dfrac{|\Gamma|}{2}=\dfrac {1}{\tau}.
\end{equation}

\subsection{Comparison with experimental data} 
This model for the quasiparticle lifetime allows us to make a prediction about the temperature dependence of the phonon frequency driven by strong anharmonicity. The expression for the phonon frequencies of Eq.\eqref{general_LOTO}, together with Eq. \eqref{klemens_summarized} and Eq.\eqref{Gamma_tau}, yields an expression for the frequency that is temperature - dependent.

\begin{figure}[h!]
    \centering
    \includegraphics[width=0.9\linewidth]{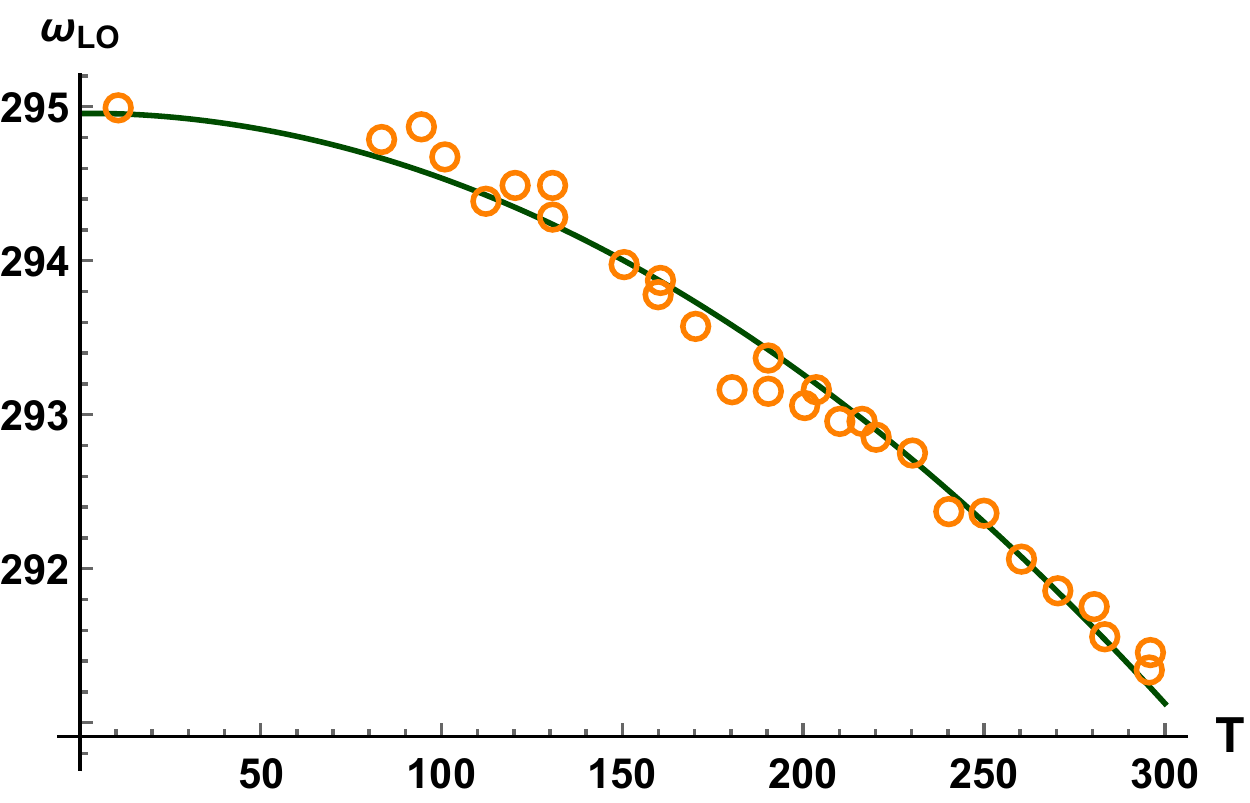}
   
    \vspace{0.2cm}
    
    \includegraphics[width=0.9\linewidth]{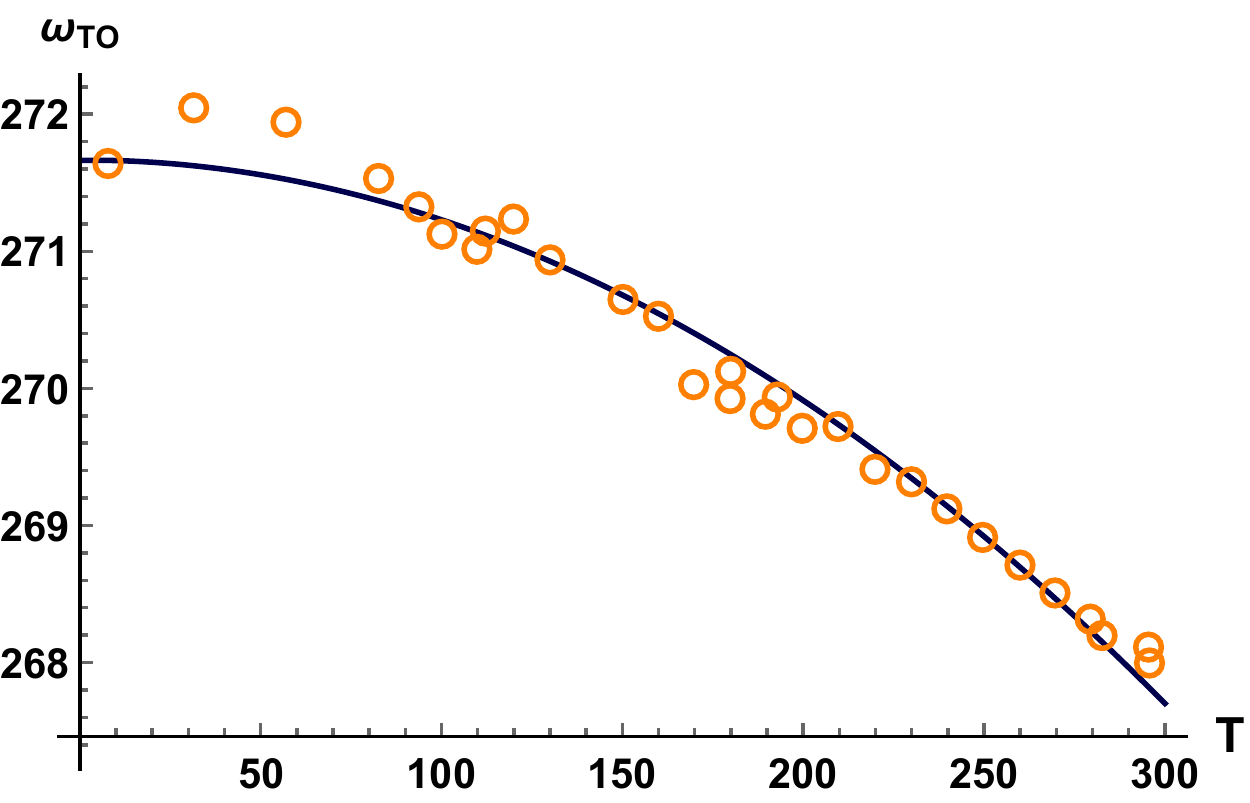}
    
    \caption{Top panel: comparison between predictions from Eq.\eqref{temperature_LOTO} (solid line) and experimental measurements (circles) of $LO$ mode frequency in GaAs as a function of $T$, obtained with the following parameters: $\omega_{LO}(0)=294.968 \,$cm$^{-1}$ and $\zeta_{LO}(0)=1.19498\times10^{-12} \,$cm$^{4}$. Bottom panel: comparison between predictions from Eq.\eqref{temperature_LOTO} (solid line) and experimental measurements (circles) of the $TO$ mode frequency in GaAs obtained with the following parameters:  $\omega_{TO}(0)=271.675 \,$cm$^{-1}$ and $\zeta_{TO}(0)=1.62005\times10^{-12} \,$cm$^{4}$. The experimental data points were taken from Ref.~\cite{Irmer}. The physical units for frequency are cm$^{-1}$, as standard for Raman measurements, while the temperature is in Kelvin.}
    \label{GaAs_TO}
\end{figure}

\begin{equation}
\label{temperature_LOTO}
  \text{Re}[\omega_{i}(T)]= \sqrt{\omega_{i}^{2}(0) -\omega_{i}^{10}(0)\zeta_{i}^{2}\left[1+\dfrac{2}{e^{x_{i}}-1}\right]^{2}}   
\end{equation}
where $i=LO,TO$. This equation is compared with experimental data in Fig. 1. It is shown that the equation provides an excellent one-parameter fitting of experimental data of temperature dependent Raman shift of softening $LO$ and $TO$ modes in model material GaAs. The original experimental data points were taken from Ref.\cite{Irmer}.

\subsection{Analytical derivation of the soft mode near $T_c$ and Curie-Weiss law}
We are now ready to study the possible emergence of a soft mode upon increasing the temperature in the ferroelectric phase. We know that a soft mode happens when the energy of a phonon goes to zero, that is the mathematical description of a mechanical instability and the indicator for a structural transition in the lattice structure. 
Imposing this condition in Eq.\eqref{general_LOTO} or Eq.\eqref{temperature_LOTO}, upon setting $\omega\equiv\text{Re}[\omega_{i}(T=0)]$, we obtain:
\begin{equation}
\label{coth}
\begin{split}
    &\omega=\omega^{5}\zeta\left[1+\dfrac{2}{e^{x}-1}\right],\\
    &\dfrac{1}{\omega^{4}\zeta}=\left[\dfrac{e^{x/2}+e^{-x/2}}{e^{x/2}-e^{-x/2}}\right]=\text{coth}\left(\dfrac{\hbar\omega}{4 k_{B}T}\right).\\
\end{split}
\end{equation}

It is easy to verify that $\text{coth}\left(\dfrac{\hbar\omega}{4 k_{B}T}\right)>1$, for any choice of temperature and of the other parameters. A simple condition for the softening of a mode in this model comes from this observation:
\begin{equation}
    \dfrac{1}{\zeta}>\omega^{4}.
\end{equation}
Whenever this condition is verified, we can derive an analytical expression for the critical temperature, $T_c$. From \eqref{coth}: 
\begin{equation}
   \begin{split}
    & \dfrac{1}{\omega^{4}\zeta}-1=\dfrac{2}{e^{x}-1},\\
    &  e^{x}= \dfrac{1+\omega^{4}\zeta}{1-\omega^{4}\zeta}.  \\
\end{split}
\end{equation}
Recalling the expression of $x$ from Eq.\eqref{klemenstau}, we obtain:
\begin{equation}
    T_{c}=\dfrac{\hbar \omega}{2k_{B} \log\left(\dfrac{1+\omega^{4}\zeta}{1-\omega^{4}\zeta} \right)}.
\end{equation}

Hence, $T_c$ is a growing function of $\omega$, so we can predict that since $\omega_{LO}>\omega_{TO}$, the critical temperature for the longitudinal soft mode will be the highest of the two.
Also, this relation and the expression for the inverse of the mean lifetime of the particle suggest that the dominant damping effect is the decay of the transverse (TO) phonon. This means that, starting from a solid with some symmetry, the first structural phase transition that we will encounter upon lowering the temperature is the one linked with the softening of the transverse phonon, in agremeent with experimental observations~\cite{Perry}.
The Klemens model thus predicts a stronger damping effect for the optical transverse phonon, which, instead, in the Barker model was assumed as the quasiparticle with infinite lifetime, which invalidates Barker's treatment discussed in the previous section.

Studying analytically this function, we find that:
\begin{equation}
\begin{split}
\dfrac{1}{\omega^{4}\zeta} \gg 1 &\Rightarrow T_{c} \rightarrow+\infty,\\ \nonumber
\dfrac{1}{\omega^{4}\zeta}-1\rightarrow 0^{+} &\Rightarrow T_{c} \rightarrow 0. \nonumber
\end{split} 
\end{equation}

Since the $\zeta$ parameter contains the Gr\"uneisen parameter, $\zeta\propto \gamma^{2}$, these limits give an insight into the effect of anharmonicity on the structural soft-mode transition.   
More precisely, for a vanishing  Gr\"uneisen parameter we have an infinitely high transition temperature $T_c$, which means that there will be no instability and no soft mode at any accessible temperature. This correctly recovers the known limit of a perfectly harmonic solid, where obviously there is no possibility of a soft mode instability. 
On the contrary, for growing values of the Gr\"uneisen parameter $\gamma$, the critical temperature $T_c$ decreases, which is physically meaningful. 
Also, the model predicts structural transitions at relatively small temperatures for solids with giant anharmonicity~\cite{NatureMaterials,NaturePhysics}.

\subsection{Behaviour near $T_c$ and Curie-Weiss law}
Let us now study the critical behaviour predicted by the model. We can obtain the critical behaviour starting from Eq.\eqref{temperature_LOTO}. First of all, we check if this expression approaches the critical point linearly. Upon taking the derivative with respect to temperature, we obtain:

\begin{equation}
\label{temperature_linear}
  \dfrac{d\text{Re}[\omega_{i}(T)]}{dT}= 
  -\dfrac{2x\omega^{10}\zeta^{2}}{\text{Re}[\omega_{i}(T)]T}\left[\dfrac{(e^{x}-1)^{2}-2}{(e^{x}-1)^{3}}\right].
\end{equation}

The limit for $T\rightarrow T_{c}$ of this expression gives:
\begin{equation}
    \lim_{T\rightarrow T_{c}} \dfrac{d\text{Re}[\omega_{i}(T)]}{dT} = -\infty.
\end{equation}

This result calls for a more careful calculation. We may start by writing the temperature as 
\begin{equation}
\label{delta}
T=T_{c}-\Delta    
\end{equation}
where $\Delta$ is a small parameter. We can expand the exponential on the r.h.s. of Eq.\eqref{temperature_LOTO}:
\begin{equation}
\begin{split}
   & \exp\left[\dfrac{\hbar\omega}{2 k_{B} (T_{c}-\Delta)}\right]=\exp\left[x_{c}\dfrac{1}{ 1-\Delta/T_{c}}\right]\\&
   \approx e^{x_{c}}\exp\left[x_{c}\dfrac{\Delta}{T_{c}} \right]\approx e^{x_{c}}\left(1+\Delta\dfrac{x_{c}}{T_{c}}\right). \\
\end{split}
\end{equation}
Similar steps lead to:
\begin{equation}
\begin{split}
       \text{Re}[\omega_{i}(T)]& \approx\sqrt{\omega^{2} -\omega^{10}\zeta^{2}\left(\dfrac{1}{\omega^{8}\zeta^{2}}-\Delta\dfrac{x_{c}(1-\omega^{8}\zeta^{2})}{T_{c} \omega^{12}\zeta^{3} }\right)}\\
       & =\left[\sqrt{\dfrac{x_{c}(1-\omega^{8}\zeta^{2})}{T_{c} \omega^{2}\zeta }}\right] \sqrt{T_{c}-T}.\\
\end{split} 
\end{equation}
This result provides, to our knowledge, the first microscopic analytical description of the $T$-dependent soft mode in the low-temperature phase of ferroelectric phase transitions. The predicted square-root cusp behaviour in Eq. (30) is in agremeent with many sets of experimental data in the literature on various materials, such as ferroelectric perovskite PbTiO$_{3}$~\cite{Scott}, ferroelectric semiconductor SbSI~\cite{Perry}, and in SrTiO$_{3}$~\cite{Fleury,Cowley1969,Yamada}.

Since the TO mode is always more damped than the LO mode, the soft mode instability will occur in the TO mode and the above derivation gives:
\begin{equation}
\omega_{TO} = \left[\sqrt{\dfrac{\hbar(1-\omega^{8}\zeta^{2})}{k_{B} T_{c}^{2} \omega\zeta }}\right]\sqrt{T_{c}-T}.
\end{equation}
Upon replacing this result in our generalized LST relation derived above, Eq. (13), and considering that the TO modes goes to zero before the LO mode, it is clear that a soft mode in the TO mode coincides with the divergence of the static dielectric constant, $\epsilon_0$.

Hence, we have derived a quantitative theory which predicts that the soft mode induced by strong anharmonic damping leads to the Curie-Weiss law for the divergence of the static dielectric constant in the low-$T$ ferroelectric phase, according to the following equation:
\begin{equation}
\frac{\epsilon_0}{\epsilon_{\infty}}=|\omega_{LO}|^{2} \abs{\dfrac{\hbar(1-\omega^{8}\zeta^{2})}{k_{B} T_{c}^{2} \omega\zeta }}^{-1}\abs{T_{c}-T}^{-1} 
\end{equation}
Importantly, this equation not only recovers the Curie-Weiss law, but also specifies the prefactors, in terms of fundamental physical constants and physical parameters. Of particular importance for material design is the dependence on the factor $\zeta$, which contains the dependence on the Gr\"uneisen parameter $\gamma_G$, on the atomic volume $a$, and on the speed of acoustic phonons $v$. This implies that the theory provides an unprecedented opportunity to tune the ferroelectric transition temperature $T_c$ by tuning the above parameters.\\

\section{Conclusion}
In summary, we started from deriving a generalized LST relation for systems with anharmonically damped optical modes. This allows us to account for the effect of anharmonic damping in the crucial TO mode that is typically associated with the soft mode instability in the low-temperature phase of ferroelectric and displacive phase transitions.
We subsequently developed a microscopic description of softening of an optical mode in the ferroelectric phase based on anharmonic damping à la Klemens, in which the leading process is the decay of the optical phonon into two acoustic phonons. Differently from the high-temperature phase where the main contribution from anharmonicity comes from the frequency-renormalization parameter $\Delta$ (consistent with Cowley's ``moderate''  anharmonicity argument), in the low-temperature phase, instead, the soft mode is driven by strong anharmonic damping of the phonon. 

The model provides a one-parameter fitting of Raman-shift softening of LO and TO modes in GaAs measured experimentally. This leads to microscopic expressions for the critical behavior of the soft mode frequency upon approaching $T_c$ from below, in ferroelectric or genereric displacive phase transitions. Furthermore, it also leads to the Curie-Weiss law in the ferroelectric phase, including physical prefactors also in this case. The analysis of the prefactors shows that both the phonon softening law as well as the critical temperature $T_c$ can be tuned via several microscopic parameters, such as the Gr\"uneisen parameter (related to the interatomic potential), the speed of sound $v$, and the atomic volume. 

These microscopic expressions provide chemical design principles for ferroelectric materials with tuneable ferroelectric transition. Furthermore, they can be used in the future in combination with the phenomenological Landau-theory framework~\cite{Littlewood,Levanyuk} to arrive at a deeper physical understanding of ferroelectrics.

\begin{acknowledgements}
AZ gratefully acknowledges financial support from US Army Research Office, contract nr. W911NF-19-2-0055.
\end{acknowledgements}


\vfill

\nocite{*}
\bibliographystyle{apsrev4-1}
\bibliography{amo}


\end{document}